\documentclass[9pt,aps,prb,twocolumn,nofootinbib,unsrtnat,superscriptaddress]{revtex4-1} 

\usepackage{amssymb,amsmath}
\usepackage[caption=off]{subfig}
\usepackage{graphicx}
\usepackage{color}
\usepackage{listings}
\usepackage{siunitx}
\usepackage{bibunits}
\usepackage{booktabs}
\usepackage{colortbl} 
\usepackage{xcolor}
\usepackage{tabularx} 
\usepackage{hyperref}
\usepackage{csquotes}  
\hypersetup{
    bookmarks=true,         
    unicode=false,          
    pdftoolbar=true,        
    pdfmenubar=true,        
    pdffitwindow=false,     
    pdfstartview={FitH},    
    pdfauthor={Blair Morrison},     
    colorlinks=true,       
    linkcolor=black,          
    citecolor=red,        
}
\usepackage[capitalise,nameinlink]{cleveref}
\graphicspath{{figures/}}

\newcolumntype{Y}{!{\vrule width -4pt}>{\centering\arraybackslash}X}
\begin{document}

\definecolor{dkgreen}{rgb}{0,0.6,0}
\definecolor{gray}{rgb}{0.5,0.5,0.5}
\definecolor{mauve}{rgb}{0.58,0,0.82}

\lstset{frame=tb,
  	language=Matlab,
  	aboveskip=3mm,
  	belowskip=3mm,
  	showstringspaces=false,
  	columns=flexible,
  	basicstyle={\small\ttfamily},
  	numbers=none,
  	numberstyle=\tiny\color{gray},
 	keywordstyle=\color{blue},
	commentstyle=\color{dkgreen},
  	stringstyle=\color{mauve},
  	breaklines=true,
  	breakatwhitespace=true
  	tabsize=3
}

\title{Compact Brillouin devices through hybrid integration on Silicon}
\author{B. Morrison} 
\author{A. Casas-Bedoya}
\affiliation{Centre for Ultrahigh Bandwidth Devices for Optical Systems (CUDOS), School of Physics, University of Sydney, Sydney, NSW 2006, Australia}
\affiliation{Australian Institute for Nanoscale Science and Technology (AINST), University of Sydney, Sydney NSW 2006, Australia}
\author{G. Ren}
\affiliation{CUDOS, School of Engineering, RMIT University, Melbourne, VIC 3001, Australia}
\author{K. Vu}
\affiliation{CUDOS, Laser Physics Centre, Research School of Physics and Engineering, The Australian National University, Canberra, ACT 2601, Australia}
\author{Y.Liu} 
\author{A. Zarifi}
\affiliation{Centre for Ultrahigh Bandwidth Devices for Optical Systems (CUDOS), School of Physics, University of Sydney, Sydney, NSW 2006, Australia}
\affiliation{Australian Institute for Nanoscale Science and Technology (AINST), University of Sydney, Sydney NSW 2006, Australia}
\author{T. G. Nguyen}
\affiliation{CUDOS, School of Engineering, RMIT University, Melbourne, VIC 3001, Australia}
\author{D-Y. Choi}
\affiliation{CUDOS, Laser Physics Centre, Research School of Physics and Engineering, The Australian National University, Canberra, ACT 2601, Australia}
\author{D. Marpaung}
\affiliation{Centre for Ultrahigh Bandwidth Devices for Optical Systems (CUDOS), School of Physics, University of Sydney, Sydney, NSW 2006, Australia}
\affiliation{Australian Institute for Nanoscale Science and Technology (AINST), University of Sydney, Sydney NSW 2006, Australia}
\author{S. Madden}
\affiliation{CUDOS, Laser Physics Centre, Research School of Physics and Engineering, The Australian National University, Canberra, ACT 2601, Australia}
\author{A. Mitchell}
\affiliation{CUDOS, School of Engineering, RMIT University, Melbourne, VIC 3001, Australia}
\author{B. J. Eggleton}
\affiliation{Centre for Ultrahigh Bandwidth Devices for Optical Systems (CUDOS), School of Physics, University of Sydney, Sydney, NSW 2006, Australia}
\affiliation{Australian Institute for Nanoscale Science and Technology (AINST), University of Sydney, Sydney NSW 2006, Australia}
\date{\today}

\begin{abstract}
A range of unique capabilities in optical and microwave signal processing have been demonstrated using stimulated Brillouin scattering. The desire to harness Brillouin scattering in mass manufacturable integrated circuits has led to a focus on silicon-based material platforms. Remarkable progress in silicon-based Brillouin waveguides has been made, but results have been hindered by nonlinear losses present at telecommunications wavelengths. Here, we report a new approach to surpass this issue through the integration of a high Brillouin gain material, $\text{As}_{2}\text{S}_{3}$, onto a silicon chip. We fabricated a compact spiral device, within a silicon circuit, achieving an order of magnitude improvement in Brillouin amplification. To establish the flexibility of this approach, we fabricated a ring resonator with free spectral range precisely matched to the Brillouin shift, enabling the first demonstration of Brillouin lasing in a silicon integrated circuit. Combining active photonic components with the SBS devices shown here will enable the creation of compact, mass manufacturable optical circuits with enhanced functionality.
\end{abstract}
\maketitle

\begin{bibunit}[apsrev4-1] 
Stimulated Brillouin Scattering (SBS) has recently emerged as an impressive tool for optical processing and radio-frequency (RF) photonics. SBS is one of the strongest nonlinearities known to optics, and is capable of providing exponential gain over narrow bandwidths of the order of tens of megahertz. This narrowband amplitude response is accompanied with a strong dispersive response, capable of tailoring the phase or group delay of a counter propagating optical signal. In light of these effects a rich body of applications have been explored such as slow light\cite{Thevenaz2008a}, stored light\cite{Zhu2007}, narrowband RF photonic filters\cite{Vidal2007,Marpaung2013c,Zhang2012}, dynamic optical gratings \cite{Song2008,Santagiustina2013}, narrowband spectrometers\cite{Domingo2005}, optical amplifiers \cite{Predehl2012,Raupach2015} and RF sources \cite{Yao1997} among others. When pumped in a resonator configuration, a narrow linewidth spectrally pure SBS laser can be generated \cite{Debut2000b,Smith1991,Geng2006}. Highly coherent lasers are used in optical communication, LIDAR and producing pure microwave sources\cite{Gross2010} among other applications. While the majority of previous works have traditionally utilised SBS in optical fiber, a number of these applications have been demonstrated in integrated form factors \cite{Pant2014a,Li2012e,Li2013c,Pant2012,Casas-bedoya2015a}. Most recently, the demonstration of \SI{52}{dB} Brillouin gain\cite{Choudhary2016a} in centimeter length scale $\text{As}_{2}\text{S}_{3}$ rib waveguides proves that performance equivalent to kilometers of optical fiber is achievable in integrated devices.

The capability to embed SBS as a functional component in active photonic circuits will enable the creation of a new class of opto-electronic devices, in particular for integrated microwave photonics\cite{Marpaung2013}. The desire to harness SBS optical processing in CMOS (Complementary metal-oxide-semiconductor) compatible platforms has recently culminated in demonstrations of SBS in various silicon on insulator (SOI) device architectures \cite{Shin2013b,VanLaer2014,Laer,Kittlaus2015b}. Underetching of different waveguide geometries is performed to create guided acoustic modes, generating strong SBS from the high opto-acoustic overlap. Initial works have demonstrated large Brillouin gain coefficients\cite{Shin2013b,VanLaer2014,Laer} in excess of \SI{e3}{m^{-1}.W^{-1}}, \num{e4} times higher than single mode optical fiber, made possible due to new interactions which occur in these subwavelength structures \cite{Rakich2012}. More recent work has focused on reducing propagation losses to improve amplification factors to more than \SI{5}{dB}\cite{Kittlaus2015b}. But in general, higher gains in SOI devices have been prevented due to nonlinear losses in silicon\cite{Wolff2015,Wolff2015a} and linewidth broadening due to small dimension fluctuations introduced during device fabrication\cite{Wolff2015b}. 
\begin{figure*}
\captionsetup[subfigure]{labelformat=empty}
 \centering
 \subfloat{\label{fig:device-schematic}}\\[-2ex]
 \subfloat{\label{fig:taper-SEM}}
 \subfloat{\label{fig:crosssection-schematic}}
 \subfloat{\label{fig:crosssection-SEM}}
 \subfloat{\label{fig:modes-neff-width}}
  \subfloat{
    \includegraphics[width=\textwidth]{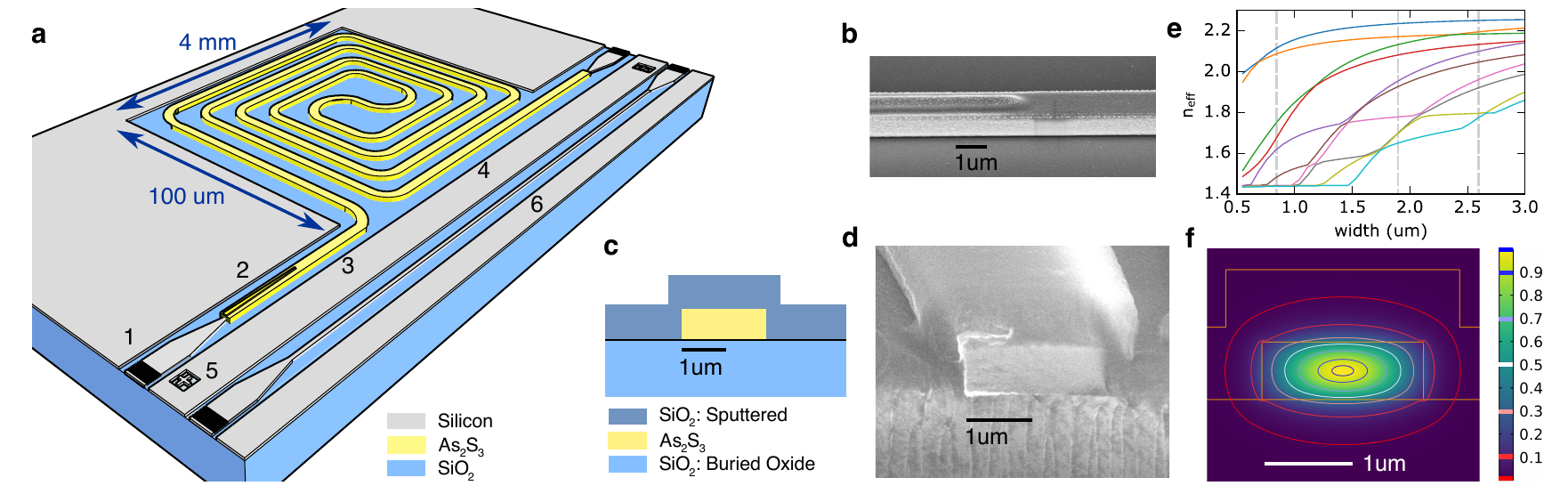}
    \label{fig:fund-mode-image}}
 \caption{\textbf{An $\text{As}_{2}\text{S}_{3}$ silicon hybrid circuit} (\textbf{a}) A schematic of the hybrid circuit with a number of components indicated 1. Silicon grating couplers with tapers to \SI{450 x 220}{nm} nanowires 2. Silicon nanowire taper region with $\text{As}_{2}\text{S}_{3}$ overlay waveguide 3. $\text{As}_{2}\text{S}_{3}$ waveguide lead into the hybrid structure 4. Spiral waveguide formed out of $\text{As}_{2}\text{S}_{3}$ 5. Alignment markers formed in the silicon layer for patterning the $\text{As}_{2}\text{S}_{3}$ structures 6. Reference silicon structures existing on the same chip (\textbf{b}) An SEM image of the end of silicon taper before cladding deposition (\textbf{c}) Schematic cross section of waveguide in chalcogenide only region (\textbf{d}) An SEM image of chalcogenide region cross section with silica cladding (\textbf{e}) Calculated effective indexes for 10 waveguide modes with increasing waveguide width. Waveguide widths used throughout the work, \SI{1.9}{um} in the spiral, \SI{2.6}{um} in the resonator and \SI{0.85}{um} in the coupler, are indicated with dashed vertical lines) (\textbf{f}) Optical mode simulation of fundamental TE mode of \SI{1.9}{um wide} $\text{As}_{2}\text{S}_{3}$ waveguide}
 \label{fig:shematic-and-stuff}
\end{figure*}%

In this work we introduce a hybrid integration approach to generate large Brillouin gain in a silicon-based device, free from nonlinear losses. We embed a compact \SI{5.8}{cm} $\text{As}_{2}\text{S}_{3}$ spiral waveguide into a silicon circuit, enabling record Brillouin gain of \SI{22}{dB} (\SI{18}{dB} net gain) on a silicon chip. Traditional silicon grating couplers are used for  coupling in and out of the chip, with silicon tapers providing low loss transitions between the Si and $\text{As}_{2}\text{S}_{3}$ sections of the circuit. To further explore the flexibility of this approach we fabricate precisely designed $\text{As}_{2}\text{S}_{3}$ ring resonators, enabling the first demonstration of Brillouin lasing in a planar integrated circuit. This work marks a significant step towards the realisation of fully integrated active SBS devices, such as integrated opto-electronic oscillators\cite{Merklein2016} and lossless microwave photonic filters\cite{Liu2016a}, in the near future.\newline 


\section*{Results}
\paragraph*{\emph{\textbf{Silicon interfaced $\text{As}_{2}\text{S}_{3}$ spiral waveguide}}}
\label{sec:Layout}
\Cref{fig:device-schematic} shows the schematic of the fabricated hybrid circuit. The circuit consists of a base silicon section and an $\text{As}_{2}\text{S}_{3}$ SBS active section. Silicon grating couplers are used for chip coupling \cite{Taillaert2006}, followed by \SI{2}{mm} long a silicon waveguide. The nanowire waveguide (\SI{450 x 220}{nm} cross section) then linearly tapers, over a length of \SI{100}{um}, to a width of \SI{150}{nm} and ends in a open silicon region of \SI{0.1x4}{mm}. Amorphous $\text{As}_{2}\text{S}_{3}$ was deposited in the open region, with a thickness of \SI{680}{nm}, completely covering the silicon tapers. Overlay waveguides were processed over the silicon taper before proceeding to the rest of the circuit, an SEM of the end of the taper region before cladding deposition is shown in \cref{fig:taper-SEM}. Optical propagation simulations, discussed in the methods, indicate a total insertion loss on the order of \SI{0.1}{dB} for transmission into the fundamental mode of the $\text{As}_{2}\text{S}_{3}$ waveguide. Silica cladding of \SI{1}{um} thickness was sputtered over the sample after $\text{As}_{2}\text{S}_{3}$ etching, with care taken to keep processing temperatures suitable for optimum losses \cite{Choi2013}. A schematic of a typical waveguide geometry is shown in \cref{fig:crosssection-schematic}, along with a cross sectional SEM \cref{fig:crosssection-SEM} and optical mode simulation of the fundamental mode of a \SI{1.9}{um} wide waveguide (\cref{fig:fund-mode-image}). Further details of the fabrication process is provided in the methods. The $\text{As}_{2}\text{S}_{3}$  region of the circuit is confined to within a small region of \SI{0.4}{mm^2} requiring significant design optimisation to achieve high performance.
\begin{figure*}
\captionsetup[subfigure]{labelformat=empty}
 \centering
 \subfloat{\label{fig:apex-pump-probe}}\\[-2ex]
 \subfloat{\label{fig:vna-spectrum}}
  \subfloat{
    \includegraphics[width=\textwidth]{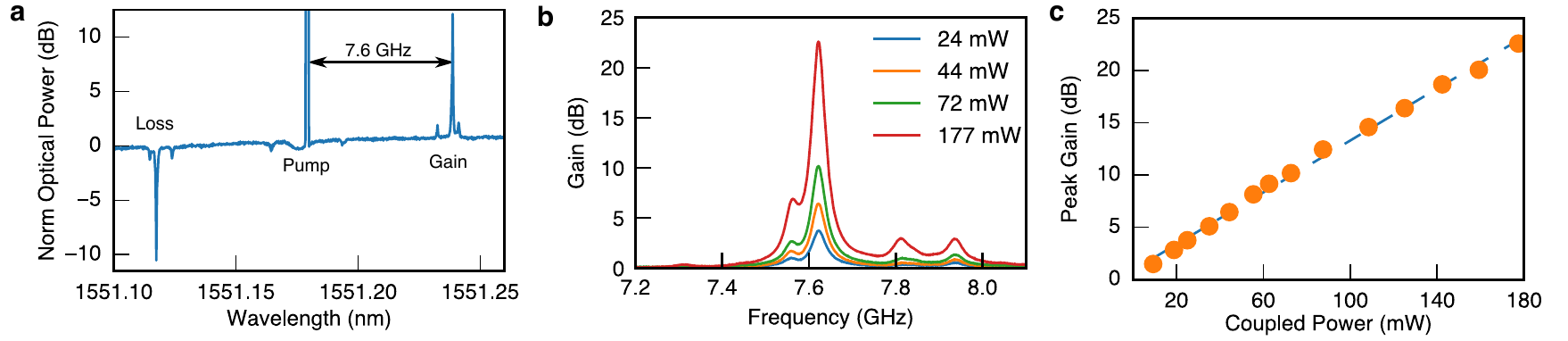}
    \label{fig:vna-peak-gain}}
 \caption{{\textbf{Backwards SBS in $\text{As}_{2}\text{S}_{3}$ spiral waveguide} (\textbf{a}) Optical spectrum measurement of SBS gain and loss. (\textbf{b}) High resolution SBS spectrum for various pump powers. (\textbf{c}) Peak gain values up to \SI{180}{mW} coupled pump power with fit.}}
 \label{fig:combined-spiral-fig}
\end{figure*}%

To maximise the physical waveguide length in the available device area (\SI{0.1x4}{mm}) we employ a folded spiral design with a rectangular shape and identical bends for each loop. The exact device geometry is chosen to give the highest Brillouin amplification. The expected gain of a weak probe, $P_{\text{o}}$, for a coupled pump power, $P_{\text{P}}$, in the small signal gain regime of backwards SBS is given by
\begin{equation}
P_{\text{S}}=P_{\text{o}}\exp(G_\text{SBS} L_{\text{eff}}\,P_{\text{P}})
\label{eq:small-sig-gain}
\end{equation}
where $G_\text{SBS}$ is the Brillouin gain coefficient and $L_{\text{eff}}$ is the effective length which is related to the physical device length $L$ by $L_\text{eff}=(1-\exp{(-\alpha L}))/\alpha$ where $\alpha$ is the linear loss. To achieve the largest gain for a given pump power we thus need to maximise $L_{\text{eff}}\times G_\text{SBS}$. The Brillouin gain coefficient is inversely proportional to the effective optical mode area, $A_{\text{eff}}$, so that the trade off becomes whether to reduce the waveguide width, $w$, to decrease $A_{\text{eff}}$ or increase the waveguide width to reduce $\alpha$, while maintaining long physical device lengths. The propagation loss is dominated by scattering losses from the rough sidewalls\cite{Poulton2006}, which has a quartic reduction with waveguide width (i.e $\alpha \propto 1/ w^4$). Larger widths lead to the waveguide becoming heavily multimoded, effective index values for the first ten guided modes are calculated for increasing widths in \cref{fig:modes-neff-width}. Adiabatic bends based on the Euler spiral\cite{Cherchi2013}, in a matched bend configuration\cite{Melloni2003}, are used in the design to minimise mode conversion, preventing extra loss throughout the structure.
$A_{\text{eff}}$ is the optical mode area
With these requirements we determined an optimum waveguide width of \SI{1.9}{um}, with effective bend radii of \SI{16.5}{um}, from FDTD simulations. A total device length of \SI{5.8}{cm} consisted of 8 loops (36 bends including external connections) with a very compact structure achieved through a small waveguide spacing of \SI{1.4}{um}. We measured a total insertion loss of \SI{4}{dB} through the spiral when correcting for the coupling losses from the grating couplers and $\text{Si}-\text{As}_{2}\text{S}_{3}$ transitions. An estimated propagation loss of \SI{0.7}{dB \per cm} resulted in an $L_\text{eff}$ of \SI{3.9}{cm} for the nonlinear interaction. The overall formfactor of this spiral represents orders of magnitude reduction compared to that of previous $\text{As}_{2}\text{S}_{3}$ waveguides used for SBS \cite{Choudhary2016a}. The typical half-etch rib geometries with multi micron widths, used for the low losses \textless\SI{0.5}{dB\per cm}, require bend radii of more than \SI{100}{um} and are incapable of high density due to the significant cross-talk introduced from the partial waveguide etch. Similarly, under etched devices require an appropriate spacing between adjacent waveguides to prevent acoustic interactions and maintain structural support, $\sim \SI{20}{um}$ width was used for a single underetched membrane structure\cite{Kittlaus2015b}.

\paragraph*{\emph{\textbf{Backwards SBS in $\text{As}_{2}\text{S}_{3}$ spiral waveguide}}}
To experimentally investigate the behavior of different devices we performed two sets of pump-probe SBS measurements, a coarse measurement using an optical spectrum analyser (OSA) and a high resolution setup with an electrical vector network analyser (VNA).
In the optical spectrum analyser measurement, a high resolution OSA (\SI{0.8}{pm}) was used to measure the transmission of a weak probe while an amplified pump laser was counter-propagated through the sample. This measurement allowed for a rough estimate of the Brillouin frequency shift in the device and enabled simultaneous monitoring of the gain and loss response. An on-off gain of \textgreater\SI{10}{dB} was observed at \SI{80}{mW} coupled power, as shown in \cref{fig:apex-pump-probe}. A Brillouin shift of $\sim$ \SI{7.6}{GHz} was measured relative to the residual back-reflected pump (centered at \SI{1551.18}{nm}) and symmetric gain and loss spectra were measured.
\newline\indent To measure the SBS response in further detail we implemented a high resolution (\textless\SI{1}{MHz}) pump-probe experiment through the use of a radio frequency vector network analyzer (VNA) \cite{Loayssa2004b}. An optical carrier undergoes single sideband with carrier modulation, using the VNA as the RF source, and is coupled into the hybrid circuit. A strong pump counter propagates through the hybrid circuit to provide gain on the modulated probe. After propagation through the hybrid circuit, the modulated optical waves beat at a highspeed photodetector and the change in RF power through the optical link is directly measured by the VNA. The photonic link response without SBS is characterised simply by turning off the pump, allowing for the backwards SBS response to be extracted as shown in \cref{fig:vna-spectrum}. Further detail is provided in the methods along with a detailed setup schematic and description in \Cref{sec:sup-pump-probe} of the supplementary. We measured the frequency spectrum for increasing pump powers up to \SI{180}{mW} coupled power, as shown in \cref{fig:vna-peak-gain}. Net amplification was achieved above \SI{25}{mW} on-chip power, overcoming the \SI{4}{dB} of propagation losses, with a maximum on-off gain of \SI{22.5}{dB} and a net gain of \SI{18.5}{dB}. This represents a $>20\times$ improvement of net gain gain compared to recent demonstrations for forward SBS \cite{Kittlaus2015b} and intermodal SBS \cite{Kittlaus2016a} in suspended silicon membrane waveguides. Fitting the slope of the measured peak gain data from \cref{fig:vna-peak-gain} obtains a Brillouin gain coefficient of $G_\text{SBS}=750\pm \SI{50}{m^{-1}.W^{-1}}$, a \SI{50}{\percent} increase over previous chalcogenide waveguides \cite{Choudhary2016a}. The linear response of the peak gain over the measured region indicates no effects due to nonlinear losses were present, further data is provided in supplementary \cref{sec:sup-nonlinear loss}. These improvements will enable a number of new applications beyond the signal processing currently demonstrated in silicon\cite{Casas-bedoya2015a}, including Brillouin lasing as explored in the next section.

\begin{figure*}
 \centering
 \subfloat{\label{fig:ring-schematic}}\\[-2ex]
 \subfloat{\label{fig:laser-concept}}
 \subfloat{\label{fig:apex-ring-measurement}}
 \subfloat{\label{fig:lasing-setup}}
 \subfloat{\label{fig:lasing-no-probe}}
 \subfloat{\label{fig:ESA-spectrum}}
  \subfloat{
    \includegraphics[width=\textwidth]{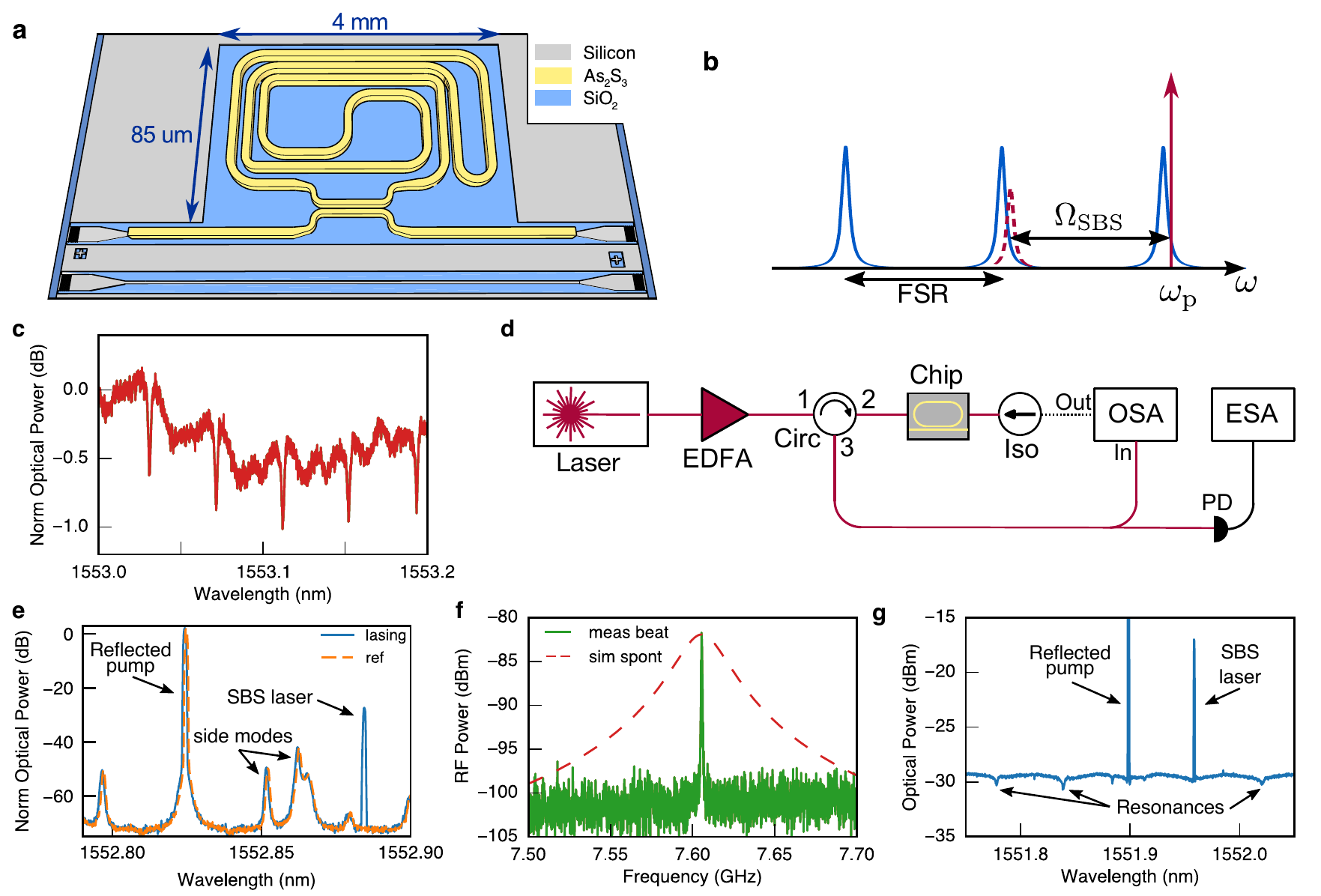}
    \label{fig:lasing-with-probe}}
 
 \caption{\textbf{Brillouin lasing in planar $\text{As}_{2}\text{S}_{3}$ resonator} (\textbf{a}) Schematic of the hybrid ring resonator structure. (\textbf{b}) A concept figure for the lasing conditions. The cavity free spectral range needs to precisely match the Brillouin shift (\textbf{c}) Typical response of ring resonator. A spacing of \SI{7.62}{GHz} and a loaded $Q$ of \num{4e5} is measured. (\textbf{d}) The setup used for measuring the laser and resonator (\textbf{e}) Lasing signal measured on OSA. The Brillouin lasing signal is observed in blue solid trace. The tunable laser is shifted slightly and the lasing no longer occurs. A number of peaks due to the modes of the laser are observable in the orange dashed trace (\textbf{f}) RF beat of the backreflected pump and lasing signal. The beat occurs at precisely \SI{7.60}{GHz}. The measured linewidth is less than \SI{5}{MHz}, significantly narrower than the natural lifetime of \SI{40}{MHz} confirming that we are above the lasing threshold (\textbf{g}) Brillouin lasing while monitoring the resonance position. Both the pump and generated Stokes are aligned to cavity resonances}
 \label{fig:full-lasing-fig}
\end{figure*}%

We explore dimensional broadening in $\text{As}_{2}\text{S}_{3}$ waveguides by measuring the SBS response of a number of different waveguide widths and lengths, including straight and spiral structures. Dimensional broadening has been identified as a key issue which reduces the expected gain, particularly in nanoscale waveguides reliant on transverse acoustic waves such as forward SBS structures \cite{Wolff2015b,Laer}. The effect manifests in changing Brillouin lineshapes as device lengths are modified, measured mechanical quality factors were reduced by almost half when moving from millimeter to centimeter scales in suspended membrane structures \cite{Kittlaus2015b}. In this work we found that the natural linewidth did not vary with device length, only minor fluctuations were measured for lengths of 1, 2, 4 and \SI{40}{mm}, results in supplementary \cref{sec:sup-broadening}. This indicates that, for these geometries, the linewidth is primarily governed by the deposited material properties of $\text{As}_{2}\text{S}_{3}$, allowing for scaling to large device lengths. \newline

\paragraph*{\emph{\textbf{Compact ring resonator}}}
Brillouin lasers are capable of spectrally narrowing laser sources and, if cascaded, can produce pure microwave frequencies. Achieving Brillouin lasing in micro resonators is challenging due to the requirement for the cavity free spectral range (FSR) to closely match the Brillouin shift. Initial demonstrations used highly overmoded resonators such that two resonances between different mode families were aligned \cite{Grudinin2009,Tomes2009}. More recently, precise matching of the cavity FSR and the SBS shift was achieved in lithographically processed silica wedge resonators\cite{Lee2012f}. These previous devices have extremely low losses, enabling low threshold oscillation, but require external coupling via tapered optical fibers or free space optics. To show a further application of the combined $\text{As}_{2}\text{S}_{3}$ and Si platform, we fabricate high $Q$ ring resonators designed for Brillouin lasing and achieve the first demonstration of Brillouin lasing in a planar integrated circuit. 

A schematic of the ring design is shown in \cref{fig:ring-schematic}. To achieve low threshold lasing in the sample we must satisfy three competing challenges; the FSR must match the Brillouin shift, the loss throughout the cavity must be minimal and the whole structure must be as compact as possible. The SBS shift scales ($\Omega_\text{SBS}$) with effective index $(n_\text{eff})$ of the optical mode and acoustic velocity of the material $(v_\text{ac})$, such that $\Omega_\text{SBS} = 2\,n_\text{eff}\,v_\text{ac}/\lambda_\text{p}$. The FSR of a resonator depends upon the total roundtrip time of the cavity, and is related to the length $(L)$ and group index $(n_\text{g})$ such that $\text{FSR}=c/(L\times n_\text{g})$. Thus we need to take into account the change of $n_\text{eff}$ and $n_\text{g}$ with waveguide width when determining the appropriate length of the resonator. The threshold for a Brillouin laser in a resonator with an FSR matching the Brillouin shift is given by \cite{Kippenberg2004a}
\begin{equation}
P_{\text{th}}=\frac{\pi^{2}n^{2}}{\lambda_{\text{p}}^{2}}\frac{L_{\text{Trip}}}{G_{B}Q_{\text{tot}}^{2}}\frac{(1+K)^{3}}{K}
\label{eq:Brillouin-threshold-ring}
\end{equation}
where $G_\text{SBS}$ is the SBS gain coefficient as above, $Q_{\text{tot}}$ is the loaded $Q$ of the resonator, $\lambda_{\text{p}}$ is the pump wavelength, $L_{\text{Trip}}$ is the roundtrip length of the resonator and $K$ is the coupling parameter which is related to the transmission $(T)$ such that $T=((1-K)/(1+K))^{2}$. To reduce propagation losses we increase the width of the waveguides up to \SI{2.6}{um}, increasing the required bend radii to \SI{22.5}{um}. Finally, to maintain a compact structure we utilised a number of individual components within the circuit. Short adiabatic couplers, based on the Milton and Burns criterion \cite{Fu2014}, were used to transition from the heavily multimoded waveguides with widths of \SI{2.6}{um} down to few mode structures with widths \SI{850}{nm}. These narrow waveguides were used in the directional coupler to provide coupling to the ring with as short length as possible. A nested spiral design, again with Euler bends, was used to minimise the footprint of the resonator and enabled the required roundtrip length ($\sim$\SI{1.5}{cm}) to fit within the required area. Further details on individual component design, including microscope images of the fabricated sample, are provided in the methods and supplementary \cref{sec:sup-simulation-microscope}.

\begin{table*}[hbt]
\centering
\caption{\bf Comparison of SBS performance in different integrated devices}
\begin{tabularx}{0.99\linewidth}{l!{\vrule width -4pt} c*{7}{Y}}  
\toprule[1pt]
Device & Type & On-Off Gain & Net Gain & $G_\text{SBS}$ & Compactness & $A_{\text{eff}}$ & FOM & NL Loss \\
{} &  & (\si{dB}) & (\si{dB}) & (\si{W^{-1}.m^{-1}}) & (\si{cm})& (\si{um^2}) & $G_\text{SBS}\times L_\text{eff}$  & Y/N  \\
\midrule[0.5pt]
$\text{Si}/\text{Si}_{3}\text{N}_{4}$ Membrane\cite{Shin2013b} & F & 0.4 & -- & 2500 & 0.5 & 0.1 & 3 & Y \\  
\rowcolor{black!20} Si nanowire pillar \cite{VanLaer2014} & F & 4.4 & -- & 3200 & 0.16 & 0.1 & 50 & Y \\ 
Si nanowire suspended\cite{Laer} & F & 2.0 & 0.5 & 6500 & 0.25 & 0.1 & 12 & Y \\ 
\rowcolor{black!20} Si membrane \cite{Kittlaus2015b} & F & 6.9 & 5 & 1150 & 3 & 0.25 & 31 & Y \\ 
Si membrane\cite{Kittlaus2016a} & $\text{F}_\text{SIMS}$ & 3.5 & 2.3 & 470 & 2.3 & 0.35 & 10 & Y \\
\rowcolor{black!20} $\text{As}_{2}\text{S}_{3}$ Rib \cite{Pant2012} & B & 22 & 16.5 & 320 & 6 & 2.3 & 13 & N \\  
$\text{As}_{2}\text{S}_{3}$ Rib spiral \cite{Choudhary2016a} & B & 52 & 40 & 500 & 2.3 & 1.5 & 40 & N \\ 
\rowcolor{black!20} This work & B & 22 & 18.5 & 750 & 0.4 & 0.9 & 30 & N \\
\bottomrule[1pt]
\end{tabularx}
\\ 
  \footnotesize{F, forward SBS; $\text{F}_\text{SIMS}$, forward stimulated intermodal scattering; B, backwards SBS; $G_\text{SBS}$, Brillouin gain coefficient; Compactness, longest dimension of SBS region; $A_{\text{eff}}$, effective mode area; NL, nonlinear; Y, Yes; N, No}
  \label{tab:sbs-device-comparison}
\end{table*}

Optical transmission measurements were performed on the fabricated device with the same high resolution OSA used in the pump-probe measurements (\cref{fig:apex-ring-measurement}). From these measurements we observed a FSR of \SI{7.62}{GHz} at \SI{1553}{nm} and an extinction ratio of around \SI{0.65}{dB} or, equivalently, a transmission of \SI{85}{\percent}, which corresponded to $K=0.04$. The measured resonance linewidth was \SI{4.5}{pm} corresponding to a $Q_\text{tot}$ of \num{4e5}. This value compares favorably to previous demonstrations of planar centimeter length scale $\text{As}_{2}\text{S}_{3}$ ring resonators with \num{3.5e5} in $\text{As}_{2}\text{S}_{3}$ on Lithium Niobate\cite{Zhou2011a} and \num{1.5e5} for directly written $\text{As}_{2}\text{S}_{3}$ waveguides\cite{Levy2015}. From \Cref{eq:Brillouin-threshold-ring} we determine an expected threshold of \SI{80}{mW} for the fabricated sample, assuming optimum matching of the SBS shift to the cavity FSR.\newline

\paragraph*{\emph{\textbf{Brillouin lasing}}}
To demonstrate Brillouin lasing in the fabricated resonator we seamlessly tuned a pump onto the resonance for a range of power levels, while monitoring the back reflected optical waves (\cref{fig:lasing-setup}). The pump light source was an external cavity laser (ECL), capable of fine resolution tuning with a continuous step size of \SI{10}{MHz}, allowing for accurate alignment to the center of the resonance. The pump was amplified before passing through a circulator (ports $1\rightarrow 2$) and coupling to the chip through silicon grating couplers. Back-scattered light from SBS, and the back-reflected pump wave, then passed through the circulator (ports $2\rightarrow 3$) and was monitored on a high resolution OSA while the RF beat was measured on an electrical spectrum analyser (ESA). A weak reference output of the OSA was also used as a probe to measure the transmission of the resonator when desired.


For coupled powers above \SI{50}{mW} Brillouin lasing was observed on the OSA and ESA. \Cref{fig:lasing-no-probe} shows a Brillouin lasing signal on the OSA, along with a reference signal at the same power level with the pump shifted just past the resonance. A single lasing signal is observed at a power level in the range of \SI{-30}{dBm}. A strong signal close to \SI{0}{dBm}, at the pump wavelength, was also measured due to the $\sim\SI{0.1}{\percent}$ backreflection of pump from the chip grating couplers. A number of cavity side modes from the backreflected pump were also observed, these are around \SI{50}{dB} below the pump signal in line with the pump laser specifications. To confirm that the measured signal was not due to spontaneous scattering we measured the electrical beatnote above threshold on the ESA (\cref{fig:ESA-spectrum}). The measured beatnote was significantly narrower than the SBS natural linewidth of \SI{40}{MHz}\cite{Choudhary2016a}, plotted with a dashed line in \cref{fig:ESA-spectrum}. The frequency of the beatnote was at $7.60 \pm\SI{0.005}{GHz}$ and slow drifts on the order of \SI{5}{MHz} were observable on the ESA over minute time scales. The lack of active locking of the pump to the resonance prevented the measurement of a slope efficiency of the Brillouin laser above the threshold level. For the weak coupling case that we have ($K=0.04$) a low slope efficiency of \SI{4}{\percent} is expected\cite{Kippenberg2004a}, improving the coupling to the ring would drastically improve this and also reduce the lasing threshold. Finally, to confirm that the Brillouin lasing is indeed occurring on the resonances of the ring, we perform an OSA measurement while sweeping a weak probe signal to measure the resonator transmission. In this case the coupled pump power was \SI{75}{mW}, \cref{fig:lasing-with-probe} shows that the lasing signal and pump are both aligned to cavity resonances.

\section*{Discussion}
To provide further details on how the $\text{As}_{2}\text{S}_{3}-\text{Si}$ hybrid circuit results compared with previous demonstrations of SBS in integrated waveguides, we prepared a comparison summary in \Cref{tab:sbs-device-comparison}. Initial silicon devices focused on achieving the highest gain coefficients possible, using highly sub wavelength structures which harnessed radiation pressure \cite{Shin2013b,VanLaer2014,Laer}. Issues arising from high scattering losses, dimensional broadening and nonlinear losses, resulted in low net gain values of below \SI{1}{dB}. More recent work has shifted to larger device geometries, resulting in interactions produced almost entirely through electrostriction. High sensitivity to local wafer conditions prevented the membrane structure from being folded, resulting in reduced compactness with straight waveguides up to \SI{3}{cm} long. However, the reduced propagation losses enabled significantly higher net gain, up to \SI{5}{dB}\cite{Kittlaus2015b}, than previous silicon demonstrations. In comparison it is clear that, being free from nonlinear losses, the $\text{As}_{2}\text{S}_{3}$ devices are capable of significantly higher on-off gain (greater than \SI{50}{dB}) compared to full silicon devices. In this work we address the compactness limitations of previous $\text{As}_{2}\text{S}_{3}$ demonstrations, through the high density and tight bends of fully etched structures, while maintaining large net gain. Finally, to understand the relative efficiency of different devices, we introduce a simple figure of merit (FOM) $G_\text{SBS}\times L_{\text{eff}}$, which is from the exponential term in \cref{eq:small-sig-gain}. To achieve \SI{20}{dB} of gain, sufficient for many microwave photonics applications\cite{Pant2014a}, with \SI{50}{mW} coupled pump power requires an FOM $\sim100$. None of the currently demonstrated devices have approached this regime, equivalent to half a km of SMF optical fiber, further improvements to the $G_\text{SBS}$ and $L_{\text{eff}}$ are expected to accomplish this goal in the near future.

One of the most desirable characteristics of Brillouin lasers is a significant linewidth narrowing of lasing Stokes lines. The key requirement to enter this regime is for the optical damping to be less than the acoustic damping or, in terms of linewidths, the cavity linewidth to be narrower than the natural linewidth of the acoustic mode\cite{Debut2000b}. If this regime is achieved then the Stokes spectrum will narrow and the fullwidth at half maximum will be given by 
\begin{equation}
\Delta v_\text{s}=\frac{\Delta v_\text{p}}{(1+\gamma_\text{A}/\Gamma_\text{c})^2}
\end{equation} where $\gamma_\text{A}$ represents the damping rate of the acoustic wave and $\Gamma_\text{c}$ is the cavity loss rate.
Thus to achieve a $100\times$ narrowing factor of the pump wave would require a cavity with $10\times$ narrower linewidth than the acoustic wave. For $\text{As}_{2}\text{S}_{3}$ waveguides with a Brillouin linewidth of \SI{40}{MHz} an optical cavity linewidth of \SI{4}{MHz} is required, corresponding to a $Q$ factor in the range of \num{5e7}. Improvements in our current fabrication processes have led to losses down to \SI{0.2}{dB\per cm} being measured in similar structures as those used in this work, leading to $Q$ factors of a few million and linewidths less than \SI{200}{MHz}. An alternative to improving the optical $Q$ is to instead reduce the acoustic lifetime, thus broadening the natural linewidth. This is possible by replacing the silica cladding with a softer cladding with acoustic velocity lower than $\text{As}_{2}\text{S}_{3}$, such as the many polymers used for lithography resists like PMMA. Simulations have shown that orders of magnitude reduction can occur in appropriate waveguide geometries \cite{Poulton2013} and experimental measurements of polymer clad $\text{As}_{2}\text{Se}_{3}$ fibers saw an increase of $10\times$ the natural linewidth\cite{Beugnot2014b}. These approaches would allow for spectral purifiers and pure microwave sources based on SBS to be implemented in fully integrated planar devices, with thresholds of a few \si{mW}, for the first time.

In conclusion, in this work we have introduced a hybrid integration approach to generate large Brillouin gain in a silicon-based device. We embedded a compact \SI{5.8}{cm} $\text{As}_{2}\text{S}_{3}$ spiral waveguide into a silicon circuit, enabling record Brillouin gain of \SI{22}{dB} (\SI{18}{dB} net gain) on a silicon chip. Precise fabrication of a compact ring resonator enabled the first demonstration of Brillouin lasing in a planar integrated circuit. Combining active photonic devices, such as modulators and detectors, with the work shown here will enable the creation of compact, high performance devices with capabilities beyond traditional RF systems.

\section*{Methods}
\textbf{Pump Probe Experiments}
A laser, frequency $\omega_{\text{c}}$, is split into two arms to create the pump and the probe wave. The pump is upshifted in frequency by $\omega_{\text{o}}$ from the carrier through the use of a Mach-Zhender intensity modulator and optical bandpass filter. The pump was then amplified with a high power EDFA, passes through ports $1 \rightarrow 2$ of an optical circulator and coupled with TE polarisation silicon grating couplers into the hybrid circuit. In the probe arm the laser undergoes single sideband with carrier modulation to produce a weak probe upshifted by frequency $\omega_{\text{RF}}$. The carrier and probe were both coupled to the device and pass through the hybrid circuit. After coupling at the output, the transmitted waves were routed from ports $2 \rightarrow 3$ of an optical circulator and then a bandpass filter was used to remove any residual back reflected pump. The remaining optical waves beat on a highspeed photodetector and the change in RF power at frequency $\omega_{\text{RF}}$ is measured by the VNA. In the frequency region of the Brillouin shift from the pump $(\omega_{\text{RF}} \approx \Omega_{\text{B}} - \omega_{\text{o}})$, the modulated sideband will experience Brillouin amplification as shown in \cref{fig:vna-spectrum}.

\textbf{Sample Preparation}
The base silicon devices were fabricated in a multi project wafer run at Imec, obtained through the europractice shuttle service (\url{http://www.europractice-ic.com}). The silicon chips were cleaned with acetone and isoproponal solution to remove protective polymer layer before thermal deposition of $\text{As}_{2}\text{S}_{3}$ \SI{680}{nm} thickness in a local region using a shadow mask process. Optical and thermal annealing stabilised the index in the range of $n=2.44$ at \SI{1550}{nm}, measured material dispersion numbers are provided in the supplementary \cref{sec:sup-simulation-microscope}. Electron beam lithography was used to pattern a spun coat ZEP resist on the $\text{As}_{2}\text{S}_{3}$. Alignment markers used for EBL were formed in the original silicon design. ICP etching was used to form the nanowires before the remaining resist was removed. Finally, sputtered $\text{SiO}_{2}$ of \SI{1}{um} thickness was deposited as a cladding layer.

\textbf{Design and Layout}
Waveguide mode simulations for calculating $n_{\text{eff}}$ and $n_{\text{g}}$, required for mask design, were performed in Comsol Multiphysics 5.0 (\url{https://www.comsol.com}), utilising measured material dispersion. The Mask design and layout of both the silicon and chalcogenide devices were performed using Ipkiss 3.1 from Luceda Photonics (\url{http://www.lucedaphotonics.com}). Custom components, Euler bends\cite{Cherchi2013} and tapers\cite{Fu2014} were defined parametrically and used throughout the circuit. Multimode chalcogenide components were optimised through 3D-FDTD simulations using FDTD Solutions 2016a from Lumerical (\url{https://www.lumerical.com}). The simulation process was automated with a number of scripts in Lumerical and Ipkiss such that generated GDSII files from Ipkiss were automatically imported into Lumerical which injected and monitored the transmission of the desired optical mode through the components. FDTD simulations incorporated material dispersion and the appropriate device cross-section. More information is provided in supplementary \cref{sec:sup-simulation-microscope}. Due to the long length of silicon tapers, 3D-BPM simulations were performed using RSoft's (\url{https://optics.synopsys.com/rsoft/}) \href{https://optics.synopsys.com/rsoft/rsoft-passive-device-beamprop.html}{BeamProp} software.

\subsubsection*{Acknowledgments}
This work has been made possible through the access to the ACT and Victorian nodes of the Australian National Fabrication Facility (ANFF). We acknowledge funding from the Australian Research Council (ARC) Centre of Excellence program (CUDOS CE110001010), Discovery grant (DP1096838) and the Laureate Fellowship (FL120100029). 
\newline
\subsubsection*{Author Contributions} 
B. J. E. and A. C.-B. conceived the device concept. G. R., K. V., D.-Y. C. and S. M. fabricated the $\text{As}_{2}\text{S}_{3}$ waveguides. T. G. N. and A. C.-B developed libraries for $\text{As}_{2}\text{S}_{3}$ layout in Ipkiss. B. M., with assistance of A. C.-B. and A. Z., performed layout for fabrication. B. M. performed simulation of $\text{As}_{2}\text{S}_{3}$ circuit components for the spiral and ring resonator. S. M. performed simulations for the $\text{As}_{2}\text{S}_{3}$ to Si taper. B. M. and A. C.-B. conducted experiments with spiral waveguides, and B. M. and Y. L. conducted experiments with ring resonator. S. M., A. M. and B. J. E. oversaw the project. B. M. wrote the paper with contributions from all authors.

\subsubsection*{Additional Information}
Competing financial interests: The authors declare no competing financial interests.

\putbib[PhDRelated-HybridPaperTex]
\end{bibunit}

\pagebreak
\clearpage
\newpage
\onecolumngrid
\begin{center}
\textbf{\large Supplemental Materials: Compact Brillouin devices through hybrid integration on Silicon}
\end{center}
\setcounter{equation}{0}
\setcounter{figure}{0}
\setcounter{table}{0}
\setcounter{page}{1}
\makeatletter
\renewcommand{\theequation}{S\arabic{equation}}
\renewcommand{\thefigure}{S\arabic{figure}}
\renewcommand{\bibnumfmt}[1]{[S#1]}
\renewcommand{\citenumfont}[1]{S#1}
\linespread{1.25}\normalsize
\begin{bibunit}[apsrev4-1]
\section{Pump Probe}
\label{sec:sup-pump-probe}

\begin{figure*}[h]
 \centering
  \subfloat{
    \includegraphics[width=\textwidth]{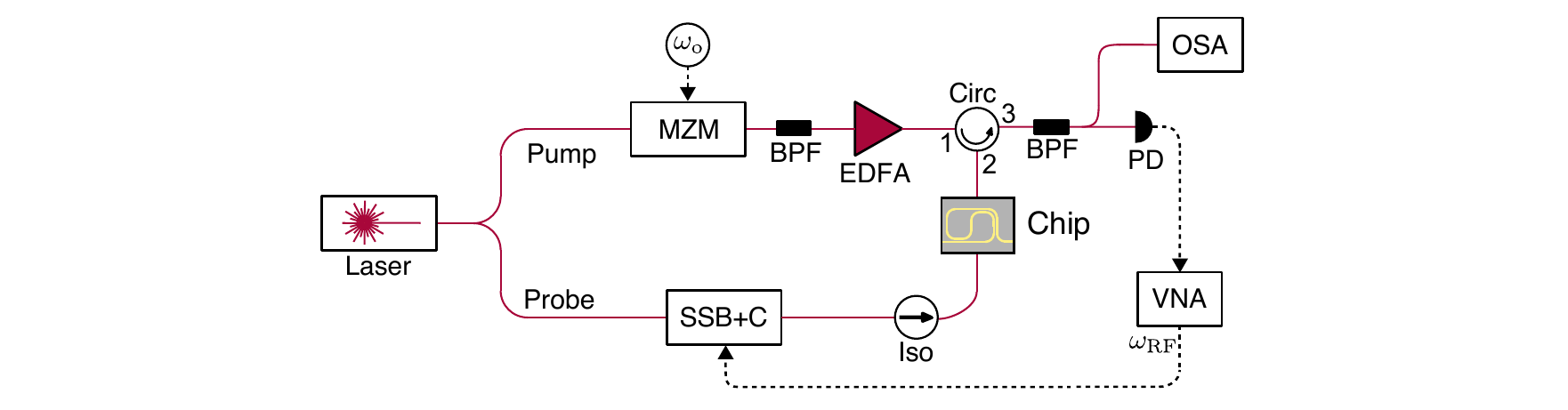}
  }
 \caption{A schematic of the simplified setup as described in the methods. The acronyms are; MZM: Mach-Zehnder modulator, SSB+C:single sideband and carrier, Iso: Isolator, BPF: Bandpass filter, EDFA: Ebium doped fiber amplifier, PD: Photodetector, OSA: Optical spectrum analyser, VNA: Vector network analyser}
 \label{fig:sup-pump-probe-schematic}
\end{figure*}

\begin{figure*}[h]
 \centering
  \subfloat{
    \includegraphics[width=\textwidth]{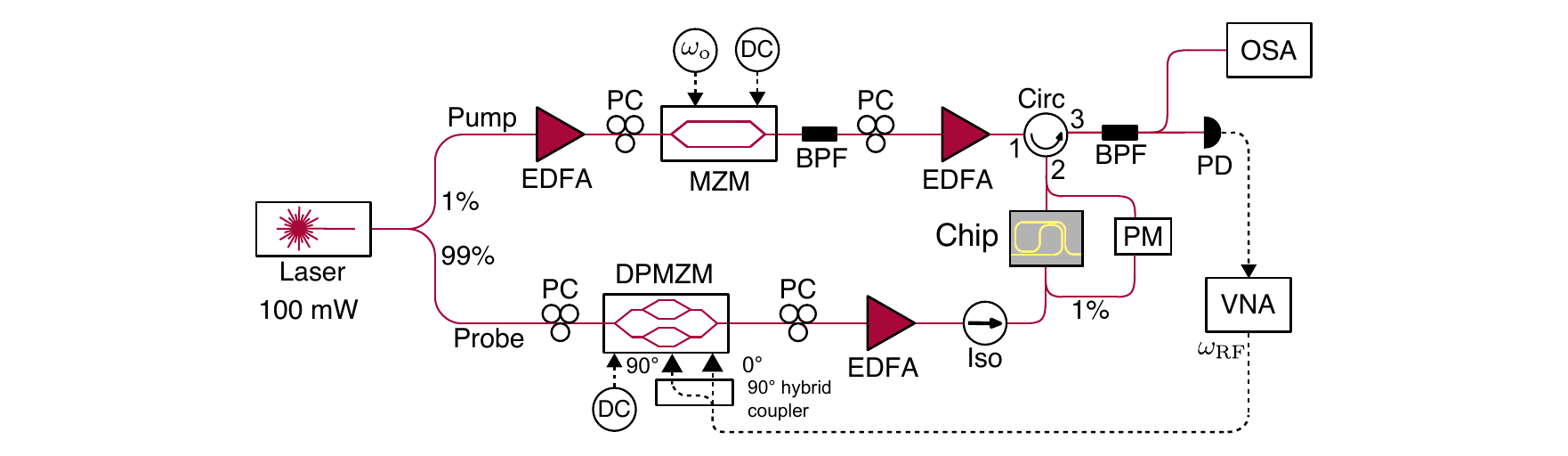}
     }
 \caption{The schematic of the full pump probe setup}
 \label{fig:sup-schematic-full}
\end{figure*}

\Cref{fig:sup-pump-probe-schematic} is the schematic of the setup which is described in the methods. Here we also provide, \cref{fig:sup-schematic-full}, a more detailed schematic outlining the overall setup and will describe its operation below.

A \SI{100}{mW} laser (with \SI{50}{KHz} linewidth) is split into two arms which form the pump and the probe. About \SI{1}{mW} enters the pump arm, is amplified with an Erbium doped fiber amplifier (EDFA) to \SI{100}{mW} and enters a Mach-Zehnder modulator (MZM). The modulator is biased with a DC voltage to suppress the original laser tone and modulated at $\omega_\text{o}$ generating two sidebands. The upper sideband is selected using a narrow $(\sim \SI{10}{GHz})$ bandpass filter and is amplified with another EDFA and sent to the chip through ports $1 \rightarrow 2$ of an optical circulator. The input to the chip is weakly tapped at \SI{1}{\%} to monitor the power, with an optical power meter (PM), and device insertion losses throughout the experiments. After transmission through the chip the remaining pump is removed with an optical isolator. In the probe arm close to \SI{100}{mW} enters a dual parallel Mach-Zehnder modulator (DPMZM). A DPMZM consists of two nested MZM and a phase shifter in one arm\cite{Kawanishi2004a}. The DPMZM is modulated at $\omega_\text{RF}$, with a hybrid coupler splitting and phase shifting the $\omega_\text{RF}$ input, from the vector network analyser (VNA) and biased with 3 DC voltages to suppress only the lower sideband (equivalent modulation can also be achieved with a dual drive MZM\cite{Smith1997}). 

Polarisation controllers (PC) are required in the setup to align to the correct polarisation axis of the modulators and the silicon grating couplers. The carrier and probe couple to the chip and will undergo SBS in the frequency region of the Brillouin shift from the pump $(\omega_{\text{RF}} \approx \Omega_{\text{B}} - \omega_{\text{o}})$. After transmission through the chip the carrier and probe pass through a bandpass filter to remove any residual back reflected pump, back reflection of the order of \SI{1}{\%} is expected from the grating coupler and cleaved SMF fiber. The carrier and probe will then beat at the highspeed photodetector (PD), producing a microwave beatnote at the original $\omega_\text{RF}$ frequency produced by the VNA. The VNA directly measures the change in RF power as $\omega_\text{RF}$ is swept in the region of interest. By turning off the pump the system response, which includes conversion from optical to microwave domain, optical transmission and also RF loss in components and cables, can be normalised and corrected directly. When the pump is turned on the SBS spectrum can be accurately extracted.

\section{Dimensional broadening}
\label{sec:sup-broadening}
\begin{figure*}[h]
 \centering
 \subfloat{\label{fig:sup-gain-narrowing-spiral}}\\[-2ex]
  \subfloat{\label{fig:sup-comparison}
    \includegraphics[width=\textwidth]{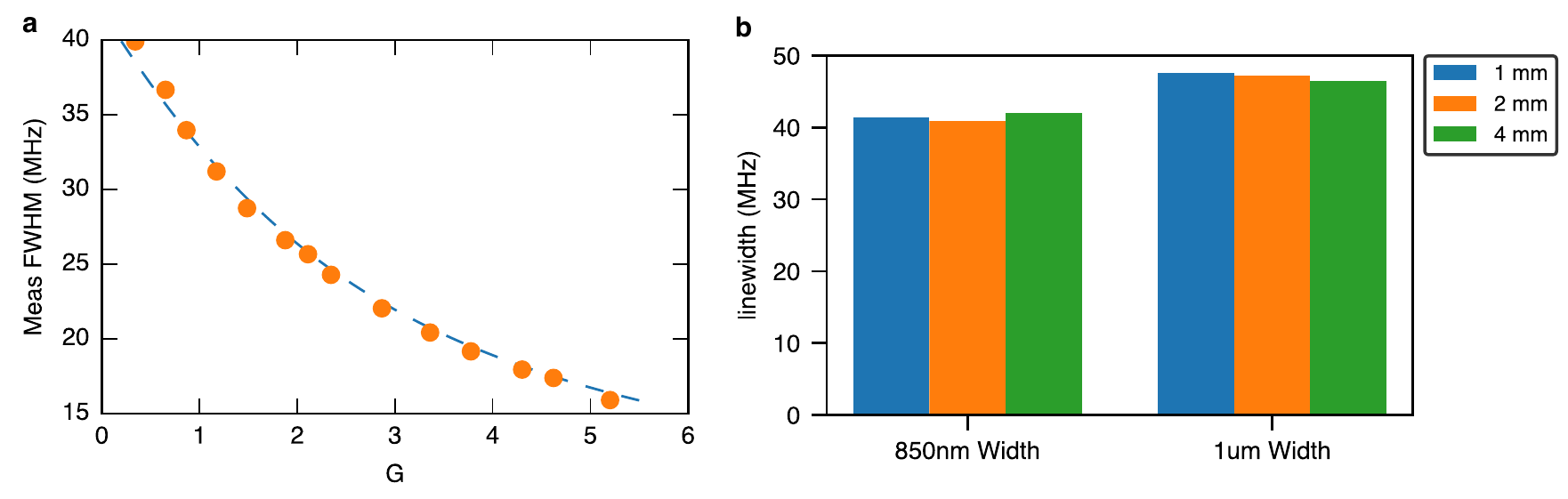}  
     }
 \caption{a) Measured linewidth narrowing with increasing gain in \SI{5.8}{cm} spiral waveguide b) Fitted natural linewidths of different waveguides}
 \label{fig:sup-broadening}
\end{figure*}

To determine if there are any issues from dimensional broadening in the $\text{As}_{2}\text{S}_{3}$ hybrid chip we measure the natural linewidth of a number of different waveguide geometries. The natural linewidth can be directly measured from the spontaneous Brillouin or inferred from pump probe measurements\cite{Yeniay2002a}. Due to the high sensitivity available with the pump probe system we used this technique for measuring the natural linewidth. The natural SBS spectrum is a Lorentzian with width $\Gamma_\text{B}$, which becomes compressed under high gain and approaches a Gaussian shape. The measured linewidth in a pump probe measurement is related to the gain through the $G$ coefficient by the following,
\begin{equation}
\Delta\nu_\text{RF} = \Gamma_\text{B}\sqrt{\frac{G}{\ln(e^G+1)-\ln 2}-1}
\label{eq:natural-fitting}
\end{equation}
where $G$ is the term in the exponential of the small signal gain equation ($G=G_\text{SBS} L_{\text{eff}}\,P_{\text{P}}$), \cref{eq:small-sig-gain} in body text. The $G$ parameter can be directly determined by the on-off gain only, making linewidth measurements robust against changes in pump power and coupling loss. Experimental measurements of linewidths with different gain values of the \SI{5.8}{cm} spiral waveguide described in the body text is shown in \cref{fig:sup-gain-narrowing-spiral}. The fitted data is also shown and the best fit yields $\Gamma_\text{B}=\SI{42}{MHz}$.

\Cref{fig:sup-comparison} shows the calculated linewidths for two sets of waveguides. The waveguides with widths of \SI{850}{nm} and \SI{1}{um} had $\Gamma_\text{B}$ of $\sim \SI{41}{MHz}$ and \SI{47}{MHz} respectively. A spiral structure (from another batch of samples) with \SI{1}{um} width and length of \SI{4}{cm} had a $\Gamma_\text{B} = \sim \SI{45}{MHz}$. The changes of natural linewidth over the lengths here are within the uncertainty range of these measurements. This is in stark contrast to forward Brillouin scattering structures which are highly sensitive to fluctuations in dimensions\cite{Laer}, a \SI{30}{\%} reduction of quality factor (directly proportional to linewidth) was observed when increasing length from \SIrange{1}{4}{mm} in silicon membranes\cite{Kittlaus2015b} (data in supplementary material). The lack of dimensional broadening with increasing length makes propagation loss the only difficulty for $\text{As}_{2}\text{S}_{3}$ devices if long SBS structures are desired.

\section{Design and Simulation Details}
\label{sec:sup-simulation-microscope}
When performing mode simulations in Comsol and propagation simulations in Lumerical the measured $\text{As}_{2}\text{S}_{3}$ refractive index was used. The measured material dispersion (for the as deposited thin film) was fit with a general Sellmeier equation in the following form with Mathematica,
\begin{equation}
n(\lambda)= B_0+\frac{B_1 \lambda^2}{\lambda^2-C_1}+\frac{B_2 \lambda^2}{\lambda^2-C_2}+\frac{B_3 \lambda^2}{\lambda^2-C_3}+\frac{B_4 \lambda^2}{\lambda^2-C_4}
\end{equation}
to allow for ease of use in Comsol. The fit values, with $\lambda$ in units of \si{um}, were 
\begin{gather*} 
[B_0, B_1, C_1, B_2, C_2, B_3, C_3, B_4, C_4] \\
\text{As}_{2}\text{S}_{3} :[3.801238, 2.229502, 0.120952, 1.205326, -1.864329, -0.148652, -1.029330 ,-1.142527, -1.743148] \\
\text{SiO}_{2} :[2.6461, 2.0750, 0.07762, 0.88840, 0.09802, 0.10101, 0.208958,0.0029796, 10.1777]
\end{gather*} 

As discussed in the body text and methods, the ring resonator and spiral waveguide required a number of individual components to be implemented in Ipkiss. The Euler bends were created in parametric fashion, following the description in Ref \cite{Cherchi2013}, using the first 5 terms of the exponential expansion for x and y components. The bending algorithm was implemented to be compatible with the rest of the Ipkiss library, enabling the bend to simply replace circular bends (or other bend shapes) where desired. This allowed the same algorithm to be used for the spiral waveguides, ring resonator and also S bends in the directional coupler. The bends were designed to operate in a matched bend configuration to maintain low losses for all bends. For the \SI{1.9}{um} width waveguides in the spiral waveguide the optimum bend radius was determined to be \SI{16.5}{um}. The wider waveguides in the ring, \SI{2.6}{um}, had a larger bending radius of \SI{22.5}{um}. 

\begin{figure*}[h]
 \centering
  \subfloat{
    \includegraphics[width=\textwidth]{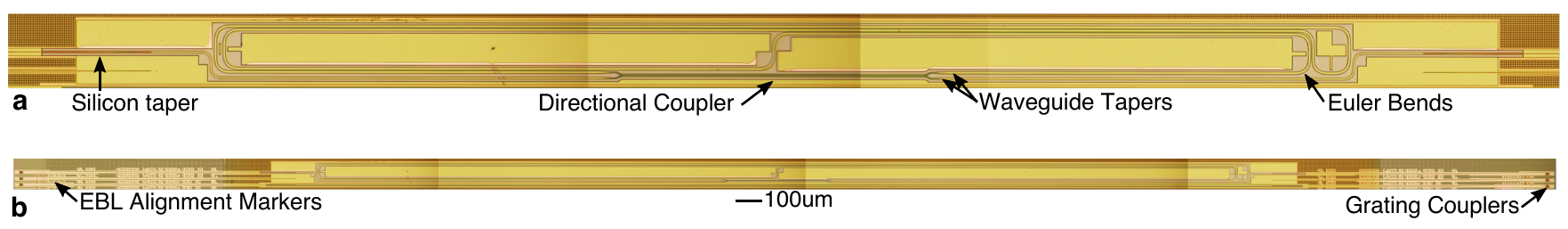}  
     }
 \caption{a) Microscope images (not to scale) highlighting key device components b) To scale microscope of entire circuit with silicon grating couplers and EBL alignment markers highlighted}
 \label{fig:sup-microscope-images}
\end{figure*}%

The significantly wider waveguides in the ring, \SI{2.6}{um}, compared to the directional coupler, \SI{0.85}{um}, required tapers for efficient mode conversion. To reduce the taper length and minimise mode conversion as much as possible, we implemented adiabatic tapers based on the Milton and Burns condition as described in Ref\cite{Fu2014}. To implement this component in Ipkiss the effective index for waveguide widths from \SI{0.5}{um} to \SI{4.0}{um} was calculated, to determine the local tapering angle for a given position. The optimised designs were determined through 3D FDTD simulations in lumerical solutions, $\alpha = 0.225$ for the $\SI{2.6}{um} \rightarrow \SI{0.85}{um}$ taper resulted in $< \SI{0.01}{dB}$ loss over a \SI{20}{um} length. For the bus waveguide, $\alpha = 0.125$ for the $\SI{1.9}{um} \rightarrow \SI{0.85}{um}$ resulted in similar losses over \SI{16.5}{um} length. The directional coupler consisted of \SI{400}{um} long coupling region with \SI{850}{nm} gap, and s bends with a radius of \SI{16}{um} and bend angle of \SI{25.0}{\degree}. A single round trip consisted of the coupler region and \SI{14.9}{mm} of \SI{2.6}{um} wide waveguide.

For consistent simulations while varying device parameters (bend radius for example) we implemented a number of scripts in Ipkiss and also Lumerical FDTD (based on scripts available with Ref\cite{Chrostowski2015a}). The scripts would import the GDS layout as created in Ipkiss, create the appropriate cross section with material dispersion for core and cladding and define the simulation region. A simulation would propagate power in a specified waveguide mode and record the relative transmitted and backreflected power in the desired (typically first 10) waveguide modes. The simulated powers were monitored in a region from \SIrange{1525}{1575}{nm}, the simulated field profile at \SI{1550}{nm} was also recorded for reference. Microscope images of the fabricated ring, with a number of components highlighted, are provided in \cref{fig:sup-microscope-images}. 



\begin{figure*}[h]
  \centering
  \includegraphics{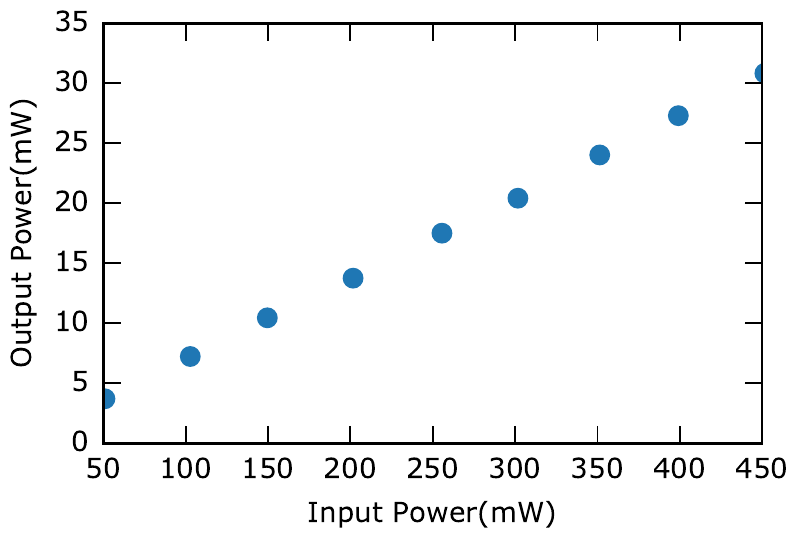}
  \caption{Transmitted power through a \SI{4}{mm} hybrid waveguide}
  \label{fig:sup-nonlinear-loss}
\end{figure*}%

\section{Nonlinear loss measurements}
\label{sec:sup-nonlinear loss}
Nonlinear losses are frequently an issue in silicon on insulator (SOI) circuits at \SI{1550}{nm}. Two photon absorption (TPA) gives rise to free carriers, which have high absorption coefficients and optical dispersion, and quench transmission through free carrier absorption (FCA). In typical SOI waveguides of dimensions \SI{450x220}{nm} the free carrier lifetime is of the order of a few nanoseconds. Complete saturation of transmitted power can occur above \SI{100}{mW} of coupled power in lengths as short as \SI{1.5}{cm}\cite{Morrison2016}. We measured transmitted powers for the majority of pump probe experiments and found nonlinear loss to be negligible. In \cref{fig:sup-nonlinear-loss} we show transmitted power measurements for input powers up to \SI{450}{mW}. The short silicon leads of \SI{2}{mm} (of varying thickness) push the threshold for nonlinear losses beyond the power handling of device couplers (in the range of \SI{800}{mW} to \SI{1}{W}) and as such, nonlinear losses are not an issue for the hybrid devices presented in this work.


\putbib[PhDRelated-HybridPaperTex]

\end{bibunit}
\end{document}